\documentclass[twocolumn]{aastex62}

\usepackage{amsmath}
\usepackage{textcomp}
\usepackage{soul}

\newcommand{\redpen}[1]{{\bf\textcolor{red}{[#1]}}}

\shorttitle{A Flaring AGN In a ULIRG candidate in Stripe 82}
\shortauthors{Prakash et al.}
\begin{document}

\title{A Flaring AGN In a ULIRG candidate in Stripe 82}

\correspondingauthor{Abhishek Prakash}
\email{aprakash@ipac.caltech.edu}

\author[0000-0003-4451-4444]{Abhishek Prakash}
\affil{IPAC, California Institute of Technology
1200 E California Boulevard, Pasadena, CA 91125, USA}

\author{Ranga Ram Chary}
\affil{IPAC, California Institute of Technology 
1200 E California Boulevard, Pasadena, CA 91125, USA}

\author{George Helou}
\affil{IPAC, California Institute of Technology
1200 E California Boulevard, Pasadena, CA 91125, USA}

\author{Andreas Faisst}
\affil{IPAC, California Institute of Technology
1200 E California Boulevard, Pasadena, CA 91125, USA}

\author{Matthew J. Graham}
\affil{California Institute of Technology,
1200 E California Boulevard, Pasadena, CA 91125, USA}

\author{Frank J. Masci}
\affil{IPAC, California Institute of Technology 
1200 E California Boulevard, Pasadena, CA 91125, USA}

\author{David L. Shupe}
\affil{IPAC, California Institute of Technology 
1200 E California Boulevard, Pasadena, CA 91125, USA}
\author{Bomee Lee}
\affil{IPAC, California Institute of Technology 
1200 E California Boulevard, Pasadena, CA 91125, USA}

\begin{abstract}

We report the discovery of a mid-infrared variable AGN which is hosted by an ultraluminous infrared galaxy (ULIRG) candidate in the Sloan Stripe 82 field. \textit{WISE} \textit{J030654.88+010833.6} is a red, extended galaxy, which we estimate to be at a photometric redshift of 0.28 $\leq$ z $\leq$ 0.31, based on its optical and near-infrared spectral energy distribution (SED). The factor of two variability over 8 years seen in the \textit{WISE} 3.4 and 4.6 $\mu$m wavelength channels is not clearly correlated with optical variability in archival data. Based on our estimation of the physical parameters of the host galaxy, \textit{J030654.88+010833.6} is possibly a composite AGN/starburst ULIRG in a phase where high star formation $\sim$ 70 M$_{\odot}$ year$^{-1}$ is occurring. Our estimate of the black hole mass to stellar mass ratio also appears to be consistent with that of broad line AGN in the local universe. The long-term variability of \textit{J030654.88+010833.6} as seen in the \textit{WISE} \textit{W1} and \textit{W2} light curves is likely due to variations in the accretion rate, with the energy being reprocessed by dust in the vicinity of the AGN. 

\end{abstract}
\keywords{Galaxy, SED, AGN, light curve--- miscellaneous --- catalogs --- surveys}
\section{Introduction} \label{sec:intro}


An active galactic nucleus (AGN) is powered by the accretion of matter onto a supermassive black hole (SMBH). Hence their occurrence rates trace the formation and growth of SMBH over cosmic time \citep{peterson_1997}. AGN can exhibit flux variation on timescales ranging from minutes to years over the entire electromagnetic spectrum \citep{fitch_1967}. The underlying physics of the variability is not clearly understood. It is suggested that on short timescales, disk instabilities play a significant role \citep{kawaguchi_1998}, while on longer timescales, the fueling of gas into the nuclear regions and regulation through feedback processes dominate \citep[e.g][]{hopkins_agn_2012}.

AGN represent an evolutionary phase for many galaxies and AGN feedback is expected to have a significant impact on the star formation of the host galaxy. The ultraviolet (UV) - optical part of the Spectral Energy Distribution (SED) of an AGN is dominated by emission from the inner accretion disk \citep[e.g.][]{shakua_sunyaev_1973}. The dusty region at larger distances from the accretion disk, often referred as the `dusty torus' absorbs the light of the accretion disk and re-emits it in the near- and mid-infrared (IR), dominating the SED at wavelengths longer than $\sim$ 1.0 $\mu$m. 
The variability of an AGN not only places size limits on the accretion disk but also traces gas inflows and outflows around the AGN \citep{shields_1978}.
Obscuration by gas and dust during the early stages of black hole growth, in
the early stages of a merger, can
suppress the UV-optical continuum of AGN, until AGN feedback becomes strong enough to clear out the gas from the central regions of the galaxy \citep{Hopkins_2005}. For a better understanding of the AGN astrophysics, a combination of observations in the time-domain and across multiple wavelengths is necessary.

Ultra-Luminous InfraRed Galaxies (ULIRG) are classically defined as galaxies with $L_{IR}=L(8-1000\,\mu m) > 10^{12}L_{\odot}$. In the local universe, these are thought to be driven by mergers between gas-rich galaxies, where the interaction triggers dust-enveloped starburst and AGN activity \citep{gao_and_solomon_2004}. One of the promising ways to disentangle the source of power (AGN vs. starburst) is using the mid-IR (MIR) and far-IR (FIR) color diagnostic \citep[e.g.][]{kirkpatrick_herschel_2013}. The MIR emission from an obscured AGN appears to be dominated by hot dust emission which exceeds the photospheric emission from low-mass and evolved stars as well as the hot dust and PAH emission from star-forming regions. 

In this paper, we focus on a MIR variable AGN \texttt{J030654.88+010833.6} (hereafter, \textit{WISE-J0301}) discovered in \textit{WISE} data in the Sloan Stripe 82 field. The extreme MIR variability found in the \textit{WISE} 3.4 and 4.6\,$\mu$m\ bands is associated with smaller variability in the
reddest optical bands.
It has a bright optical, MIR, radio, X-ray and a faint ultraviolet source associated with it. SED fitting to the UV-near IR (NIR) photometry  results in a photometric redshift of 0.28 $\leq$ z $\leq$ 0.31. 

Using the template SED of Mrk231 \citep{sanders_mrk0231_1988} to fit the MIR photometry, we identify the \textit{WISE-J0301} host galaxy as a ULIRG. 
MIR variable AGN are extremely rare. To put it in context, \citet{Assef_agn_cat_2018} find 162 MIR variable AGN, with detection by FIRST at 1.4GHz (over $\sim$10,000 deg$^2$), out of a \textit{WISE} catalog of 4.5 million AGN candidates. \textit{WISE-J0301} is one of the few objects of this type (MIR variability with radio detection showing no optical variability) found in the past and provides a rare opportunity to put important constraints on the growth of SMBHs and their connection to galaxy evolution at late cosmic times. 

This paper is organized as follows. Section~\ref{sec:data} describes the source selection and multi-wavelength photometry of \textit{WISE-J0301}, ranging from UV to radio from different broadband photometric surveys. Section~\ref{sec:variability} summarizes the process leading to optical and infrared light curves. Using these light curves we demonstrate variability in the infrared brightness and in the 3.4 to 4.6 $\mu$m spectral index. Section~\ref{sec:sed_fitting} presents the best fit SED based on observed photometry in order to derive the physical parameters of the host galaxy. We present our interpretation in Section~\ref{sec:discussion} and conclude with the results in Section~\ref{sec:results}.

Unless stated otherwise, all magnitudes and flux densities in this paper are expressed in the AB system \citep{oke_gunn_1983}. We use a standard $\Lambda$CDM cosmology with $H_{0}$=70 km s$^{-1}$ Mpc$^{-1}$, $\Omega_{\rm M} = 0.3$, and $\Omega_\Lambda = 0.7$, which is broadly consistent with the recent results from {\em Planck \citep{Planck14}}.

\section{Source selection and its multi-wavelength properties} \label{sec:data}
In the course of a multi-wavelength search for MIR variable AGN in the 
270\,deg$^{2}$ Sloan Digital Sky Survey (SDSS) Stripe 82 field, we selected potential AGN candidates using the \textit{WISE} all-sky catalog and the color selection criterion outlined in \citet{stern_agn_2012}. 
The \textit{WISE} \textit{W1} and \textit{W2} bands are centered at 3.4 and 4.6\,$\mu$m, respectively. $\textit{W1 - W2} > 0.8$ mag (Vega) identifies $\sim$51 AGN candidates per square degree to a depth of \textit{W2} = 15.0\,mag (Vega). We then construct the light curves for each AGN candidate using \textit{WISE} single exposure images which is explained in detail in Section~\ref{sec:variability}.

\textit{WISE-J0301} stands out among 300 AGN, whose light curves were visually inspected, with what appears to be an extreme MIR variability of $\sim$0.6 magnitudes which roughly corresponds to a flux density ratio of 1.7. The variability is seen over a time frame of 8 years (with a gap of 3.5 years after the WISE all-sky survey) and
is correlated in both \textit{W1} and \textit{W2} filters of \textit{WISE}. In addition, we find four other MIR variable AGN, with a variability of $\lesssim$ 0.4 magnitudes

\textit{WISE-J0301} has an (RA, Dec) of (46.728676$\arcdeg$, 1.142683$\arcdeg$)
and has high signal to noise detections across different surveys ranging from UV wavelengths to radio wavelengths. SDSS DR14 \citep{SDSS_DR14} and Pan-STARRS photometry \citep{PS_overview_2016} for \textit{WISE-J0301} are broadly consistent with each other (see Table~\ref{tab:phot_table}). \textit{WISE-J0301} appears to be a red extended galaxy in these imaging datasets (Figure~\ref{fig:ps_wise_img}). 
We also obtain UV photometry from the GALEX point source catalog provided by \citet{GALEX_overview_2005}. Infrared photometry is obtained from WISE All-Sky Release \citep{WISE_overview_2010} and 2MASS point source catalog \citep{2MASS_overview_2006} search service provided by the Infrared Science Archive (IRSA)\footnote{\url{http://irsa.ipac.caltech.edu/}}. \textit{WISE-J0301} also has a radio detection in NVSS 1.4GHz band \citep{NVSS_overview_1998}. These photometric measurements are summarized in Table~\ref{tab:phot_table} below.

\begin{figure}[ht!]
\plotone{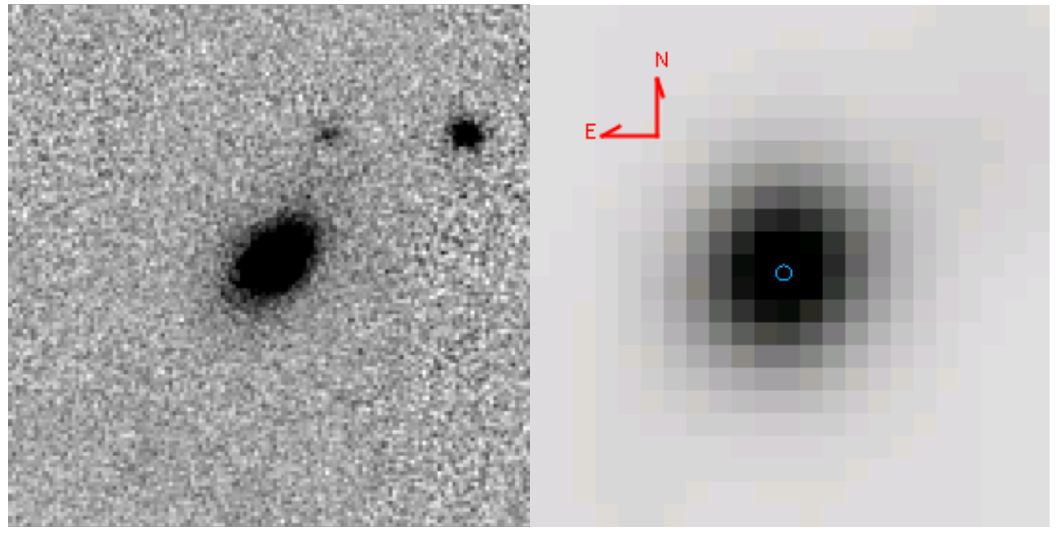}
\caption{35 arcsec across Pan-STARRS i-band (left panel) and \textit{WISE} \textit{W1}(right panel) images of \textit{WISE-J0301}. The host galaxy appears to be a red extended galaxy.}
\end{figure}
\label{fig:ps_wise_img}

In order to understand better the nature of the MIR variability, we also obtain multi-epoch optical photometry from the Catalina Real-Time Transient Survey (CRTS, \citealt{drake_catalina_2009}), and PanSTARRS surveys \citep{PS1_2016}. 


\begin{table}[ht]
\tabletypesize{\scriptsize}
\centering
\begin{tabular}{llll}
\hline
\hline
Filter Name & $\lambda(\mu m)$  & Mag$_{AB}$/Flux(mJy) \\
\hline
\hline
\textit{GALEX FUV} & 0.15 & 22.87$\pm$0.34\\
\textit{GALEX NUV} & 0.23 & 23.40$\pm$0.50\\
SDSS u$_{Model}$ & 0.36 & 21.27$\pm$0.18 \\
SDSS g$_{Model}$  & 0.47 & 19.39$\pm$0.02\\
SDSS r$_{Model}$  & 0.62 & 18.05$\pm$0.01\\
SDSS i$_{Model}$  & 0.75 & 17.53$\pm$0.01\\
SDSS z$_{Model}$  & 0.88 & 17.20$\pm$0.01\\
PanSTARRS g$_{Kron}$  & 0.47 & 19.45$\pm$0.02 \\
PanSTARRS r$_{Kron}$  & 0.62 & 18.33$\pm$0.01 \\
PanSTARRS i$_{Kron}$  & 0.75 & 17.78$\pm$0.01\\
PanSTARRS z$_{Kron}$  & 0.88 & 17.59$\pm$0.02\\
PanSTARRS y$_{Kron}$  & 0.96 & 17.40$\pm$0.02\\
2MASS J & 1.23 & 17.25$\pm$0.11\\
2MASS H & 1.66 & 16.95$\pm$0.09\\
2MASS K$_s$ & 2.16& 16.33$\pm$0.06  \\
\textit{WISE W1} & 3.4 & 15.92$\pm$0.02\\
\textit{WISE W2} & 4.6 & 15.36$\pm$0.02\\
\textit{WISE W3} & 12.0 & 14.36$\pm$0.04\\
\textit{WISE W4} & 22.0 & 12.56$\pm$0.11\\
\textit{Spitzer IRAC ch-1} & 3.6 & 15.48$\pm$0.002\\
\textit{Spitzer IRAC ch-2} & 4.5 & 15.17$\pm$0.001\\
NVSS 1.4GHz & 2.14$\times 10^5$ & 14.65$\pm$0.08 / 5.0$\pm$0.04\\
\hline
\end{tabular}
\caption{\textit{WISE-J0301} time-averaged multi-wavelength photometry obtained from different imaging surveys. All magnitudes are in the AB system.}
\label{tab:phot_table}
\end{table}

\section{Variability in the infrared and optical light}\label{sec:variability}
\subsection{MIR variability}\label{sec:infra_lc}
The infrared light curves in \textit{WISE} \textit{W1} and \textit{W2} bands, are generated for each AGN candidate by applying standard aperture photometry on publicly available \textit{WISE} single exposure (Level 1b) images acquired from IRSA image service\footnote{\url{http://irsa.ipac.caltech.edu/applications/wise/}}. The Single-exposure \textit{WISE} images contain a large and diverse quantity of image artifacts. To overcome these, we restrict ourself to good quality images and apply the following quality control criteria as suggested by the \textit{WISE} team\footnote{\url{http://wise2.ipac.caltech.edu/docs/release/allsky/expsup/sec2_4b.html}}:
\begin{eqnarray}
\texttt{qual\_frame = 10}, \label{eqn:img_qual}\\
\texttt{SAA\_SEP > 0, and} \label{eqn:saa}\\
\texttt{MOON\_SEP > 24}. \label{eqn:moon}
\end{eqnarray}
Equation~\ref{eqn:img_qual} is a good overall quality score. This parameter is available in the Single-exposure (level 1b) Frame Metadata table and can assume the values of 0, 5, and 10. The images with $qual\_frame < 10$ are poor-quality frames and not considered science worthy. These images may be affected by a combination of artifacts, included but not limited to, high pixel noise, high background, and degraded image quality due to spacecraft jitter. 
Equation~\ref{eqn:saa} ensures that the images were taken when the 
spacecraft was well outside the nominal boundaries of \textit{South Atlantic Anomaly} and Equation~\ref{eqn:moon} selects images that are not affected by the worst of moonlight contamination. Almost 25-30$\%$ of images which fail to meet the standards stated above are excluded from our analysis.\\ 
\indent Each \textit{WISE} single exposure image is accompanied by a corresponding bit-wise encoded mask image. We use the bitwise value assigned to each pixel in an image to mask out bad pixels.  A bad pixel can be affected by one or more systematic artifacts like, saturation, dead pixel, high dark current amongst others. However, we ensure that there are no bad pixels within the photometry extraction aperture.
More detail on this can be found in the \textit{WISE} data product explanatory supplement.\footnote{\url{{http://wise2.ipac.caltech.edu/docs/release/allsky/expsup/sec4\_4a.html\#maskdef}}} We apply a circular aperture with radius fixed at 6 arc-seconds at AGN positions to estimate flux densities. To estimate the sky background and photometric uncertainty, we use a circular annulus around the source position with radii 3-3.5 times the aperture radius. A sigma clipped median provides a robust estimate for background levels while accounting for potential contamination of the background value by unrelated sources. An aperture radius of 6$\arcsec$, which is the spatial resolution (full width at half maximum) in \textit{WISE} \textit{W1} and \textit{W2} channels, helps us avoid contamination from nearby sources. However, this results in a loss of estimated flux for which we apply aperture correction using standard stars. 
The aperture corrections are derived by comparing
our aperture fluxes with the  source catalog \textit{W1MPRO} fluxes for a large number of stars ($\sim$50). We estimate aperture corrections of 0.26 and 0.31 mags in \textit{WISE} \textit{W1} and \textit{W2} channels, respectively. Our estimates for aperture corrections are consistent with similar estimates made by \textit{WISE} team.\footnote{\url{{http://wise2.ipac.caltech.edu/docs/release/allsky/expsup/sec4\_4c.html\#circ}}} Although aperture corrections are slightly sensitive to the source spectral index across the bandpass, with sources having redder spectra having larger corrections, this correction is small and therefore ignored. Also, even though the source is extended in the ground-based optical data, it is only marginally resolved at WISE resolution and so the
aperture corrections for a 6$\arcsec$ radius aperture derived through the reference stars should be reasonably accurate.
We use \texttt{Photutils} \citep{photutils} package distributed with \texttt{Astropy} \citep{astropy_2013} python module for employing aperture photometry.

We present the light curves for \textit{WISE-J0301} generated from \textit{WISE} \textit{W1} and \textit{W2} single exposure images in the top two panels of Figure~\ref{fig:wise_lc}. \textit{WISE} magnitudes are normally expressed in the Vega system. Here, we have converted the magnitudes to the AB system.\footnote{W1$_{AB}$ = W1MPRO(Vega) + 2.699, W2$_{AB}$ = W2MPRO(Vega) + 3.339} For generating these light curves, we have combined data from all different periods of \textit{WISE} observations including \textit{NEOWISE} \citep{neowise}. The $\sim$3.5 years of gap in \textit{WISE} light curves around MJD 55725 come from the period after the all-sky survey when the telescope was in hibernation. We see a clear evidence of variability of $\sim$0.6 - 0.7 magnitudes which are correlated in both the MIR channels of \textit{WISE}. Using the SED modeling described in the next section, we estimate the contributions to the photometry from
the stars within the host galaxy in \textit{WISE} \textit{W1} and \textit{W2} filters to be 370 and 260 $\mu$Jy, respectively. Subtracting these flux densities from the light curves enables us to analyze the variability of only the AGN. The photometric uncertainties in single epoch magnitudes are $\leq$0.03 mag.

The last data point in each of the panels in Figure~\ref{fig:wise_lc} (except panel 3 \redpen{and 4}) is from the first epoch of our observations of \textit{WISE-J0301} using \textit{Spitzer IRAC} channel 1 $\&$ 2 (corresponding to 3.6 and 4.5$\mu$m). We employ aperture photometry on \textit{Spitzer} post-BCD (basic calibrated data) science level images using the same procedure applied on \textit{WISE} images. We use an aperture of radius 6 arc-seconds to estimate flux densities and a sigma clipped median in the circular annulus to estimate the background flux. We then apply an aperture correction of 1.047, corresponding to the aperture radius, on the fluxes obtained from aperture photometry-we have similar caveats for this aperture correction as noted above for WISE data. We estimate the spectral index ($\alpha$) using the fluxes in \textit{Spitzer IRAC} ch-1 $\&$ 2. This spectral index is used to determine the color corrections that need to be applied to the \textit{Spitzer IRAC} fluxes. Having derived the aperture and color-corrected final flux densities in \textit{Spitzer IRAC} ch-1 $\&$ 2, we re-estimate the spectral index. This spectral index in combination with \textit{Spitzer} flux densities is used to estimate the flux densities in \textit{WISE} \text{W1} and \textit{W2} channels using the power-law in equation~\ref{eqn:p_law_wise}. Details on aperture corrections and color-corrections applied here are explained in the \textit{Spitzer IRAC} instrument handbook.\footnote{\url{{https://irsa.ipac.caltech.edu/data/SPITZER/docs/irac/iracinstrumenthandbook/}}} 

\begin{figure*}[ht!]
\includegraphics[trim=0.0in 0.0in 0.in 0.in, clip=true, width=1.0
\textwidth]{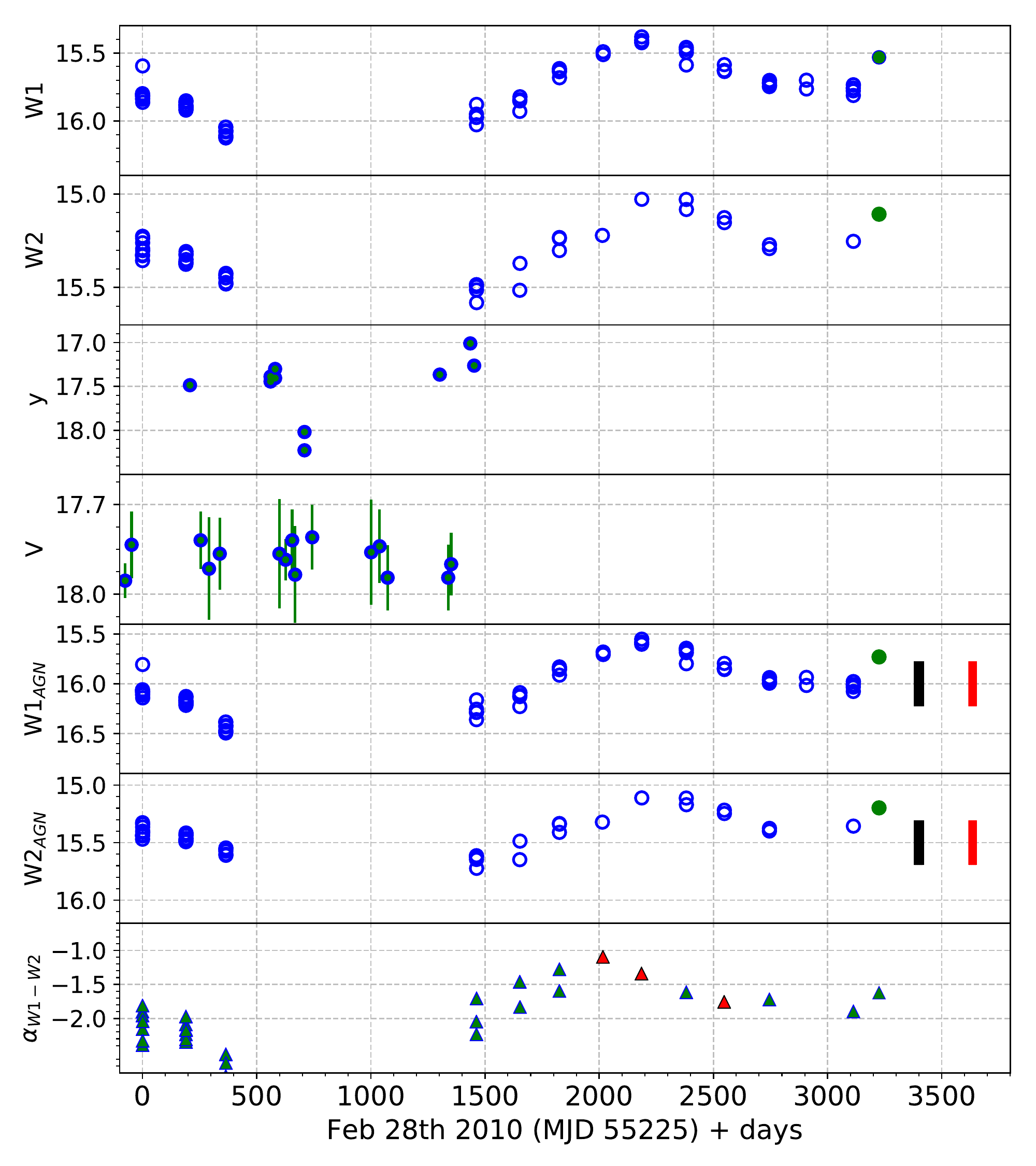}
\caption{The light curve of AGN \textit{WISE-J0301} in \textit{WISE} \textit{W1}, \textit{W2}, PanSTARRS $y-$band and CRTS V bands. The gap in \textit{WISE} data between 400 to 1400 days is the period when \textit{WISE} satellite was in hibernation post All-WISE survey. Panel 5 and 6 show the corresponding WISE light-curves after removing the stellar contributions to these wavelengths which are estimated using SED fitting. The last data point, in each panel, is the first epoch of observation from \textit{Spitzer IRAC}. The two bars in black, and red are the two additional epochs where \textit{Spitzer IRAC} will observe \textit{WISE-J0301} at 3.6, and 4.5 $\mu$m via an approved DDT program. The bottom panel shows the spectral index calculated from \textit{WISE} \textit{W1} and \textit{W2} bands after removing the stellar contribution to the flux. The red spectral index points were from data taken within a day of each other. The photometric uncertainties in single epoch \textit{WISE} and PanSTARRS magnitudes are $\leq$0.03 mags.}
\end{figure*}
\label{fig:wise_lc}

The bottom panel of the Figure~\ref{fig:wise_lc} shows the \textit{W1-W2} spectral index as a function of time using a simple power law:
\begin{eqnarray}
F_{\nu} & \propto & \nu^{\alpha}, \label{eqn:p_law}\\
log_{10}\left(\frac{F_{W1}}{F_{W2}}\right)  & = &\alpha \times log_{10}\left(\frac{\nu_{W1}}{\nu_{W2}}\right), \label{eqn:p_law_wise}
\end{eqnarray}
For spectral index, we only use the data on \textit{WISE} \textit{W1} and \textit{W2} which have the same observation date. We see the variability in the spectral index between $-2.7 < \alpha < -1.2$. A steep negative spectral index typically indicates synchrotron radiation although a significant contribution from hot dust and/or the redshifted 3.3$\mu$m polycyclic aromatic hydrocarbon feature into the WISE 4.5$\mu$m bandpass cannot be ruled out \citep{DraineLi2007}. For example, we estimate using the \citet{DraineLi2007} models that
the hot dust continuum can have an $\alpha\sim -0.4$ while the 3.3$\mu$m PAH can result in an 
$\alpha\sim -6$. Thus, varying the ratio of hot dust to PAH can account for the range of alpha
that we observe.

\subsection{Optical variability}\label{sec:opt_lc}
Once we had identified \textit{WISE-J0301} in the MIR light curves, we investigated this variability further using the optical light curve taken from publicly available data of CRTS \citep{drake_catalina_2009} shown in the fourth panel of Figure~\ref{fig:wise_lc}. CRTS data is taken in an unfiltered optical light which is later calibrated to a Johnson V-band ($\lambda \sim$ 0.55 $\mu$m) zero-point. CRTS has 340 observations on \textit{WISE-J0301} with a baseline of 8 years between September 14, 2005 to Oct 26, 2013.\footnote{\url{{https://heasarc.gsfc.nasa.gov/cgi-bin/Tools/xTime/xTime.pl}}} Here, we group the data together in similar time-bins as the \textit{WISE} data and use the median value to provide better visibility of the trend. The gaps in the optical light curve are inherent in the CRTS data and not a consequence of grouping the data. We do not see any clear evidence of variability in the optical light curve. However, a variability of $\sim$0.3 mag or less would be hard to detect due to the sensitivity and larger photometric uncertainties of CRTS. 

A comparison between the time-averaged SDSS and PanSTARRS photometry in Table 1 shows very
significant variability in the redder optical bands, even after accounting for differences in how the photometry is performed. Specifically, we see variability of 0.39 mags in $z$, 0.25 mags in $i$, 0.28 mags in $r$, which is smaller than what we see in the WISE bands. PanSTARRS DR2 also provides 10 epochs of photometry between MJD 55433 and 57000 overlapping with the early part of the WISE light curve as shown in Figure 2. Those data show clear evidence of a factor of 3 variability with a minimum in the $y-$band flux density at about MJD of 56000. This mostly corresponds with the gap in the WISE light curve and therefore it is challenging to use this to estimate a time delay between the optical and mid-infrared variability. The absence of variability in the CRTS optical data is therefore because the spectral index of the emission is steep which results in most of the emission being in the redder bands. Thus, for this source at least, the CRTS data are insensitive to the variability.

The red colors of the source in the optical and strong MIR emission suggests that the light at longer wavelengths may be dominated by hot dust emission in the vicinity of the AGN.  In canonical accretion disk models, the short wavelength emission arises from the hot accretion disk around the black hole which is then thermally reprocessed by dust at larger distances \citep[e.g.][]{peterson_1997}. In this particular case, the spectral index of just the time-varying component of optical emission is rather steep; the spectral index of the time varying component of emission between the $i-$ and $z-$bands 
is $\alpha\sim -4$, steeper than but similar to that in the WISE bands. 
Furthermore, the absence of similarly strong variability in the shortest
wavelength optical band (e.g. $g$) argues in favor of dust extinction internal to the host galaxy being responsible for suppressing that emission since extinction is a factor of $\sim$3 higher in the $g-$band than in the $y-$band. We also calculate that it is not possible for the entire optical through MIR emission to arise from an extincted accretion disk since it would require $\sim$30 mags of visual extinction, implying a much steeper optical slope than observed and a correction to the X-ray luminosity of a factor of 40, which appears inconsistent with the correction estimated in Section 5.4.

In the next section, we use the multi-wavelength photometry summarized in Table~\ref{tab:phot_table} to model the SED of the host galaxy and derive physical parameters.

\section{SED Modeling: Deriving Physical Parameters} \label{sec:sed_fitting}
In order to understand the nature of the galaxy that is hosting such an AGN, we derive the properties of the stellar population using the high signal to noise photometry obtained from different surveys ranging from UV-radio wavelengths (Table~\ref{tab:phot_table}) with an exception of FIR\footnote{We searched both the Akari and Herschel archives and did not find any detections in the former or data in the case of the latter.}.

We fit Flexible Stellar Population Synthesis (FSPS) models \citep{fsps_2009_conroy} to
the UV-NIR SED using the EAZY package \citep{eazy} 
to derive a redshift and physical properties of \textit{WISE-J0301}. 
EAZY uses a linear combination of 12 galaxy templates derived from FSPS models. This method was originally implemented for a different set of templates within EAZY \citep{eazy} but has since been modified to use the FSPS templates.

While SED fitting, we exclude MIR photometry to avoid bands where the AGN and/or dust emission may be dominating. We estimate a photometric redshift (photo-z) with 9 bands of GALEX+SDSS+2MASS, and 10 bands of GALEX+SDSS u+Pan-STARRS+2MASS, separately, as shown in Table~\ref{tab:photo_z_table}. Independent estimates of redshift using the Sloan and Pan-STARRS are slightly different from each other, $z=0.31$ with SDSS and $z=0.28$ with Pan-STARRS, although Pan-STARRS provides a better $\chi^2$ statistics. The differences between the two observations are mainly because optical photometry are taken at different time epochs and are variable. Moreover, the photo-z estimate with Pan-STARRS benefits from using more bands with an additional y-band and thus has smaller $\chi^2$ value.

In addition, since there is an evidence for AGN contributions in the red optical bands, we also fit the SED using 24 AGN and hybrid templates (AGN+host galaxy) of  \citet{Salvato_AGN_temp_2009ApJ}. The Pan-STARRS photometry is best fitted by the hybrid template (S0-70\_QSO2-30) constructed using 70\% S0 galaxy and 30\% type 2 QSO and the derived photo-z is $z=0.28$, identical to the one derived from the FSPS galaxy templates. The $\chi^2$ value however is worse with AGN templates (8.48 vs. 3.23) because of a lack of flexibility of the SED fitting using a limited number of templates. When fitting to the SDSS photometry with these templates, the fits prefer a different hybrid template (S0-10\_QSO2-90) and results in $z=0.47$ with a significantly larger $\chi^2$ ($=18.35$) value than Pan-STARRS, indicating a poor fit.

Even though the $\chi^2$ values using SDSS data are worse when using both galaxy and AGN templates, we cannot conclusively rule out different photo-z estimates obtained from SDSS and Pan-STARRS based on the fact that the multiwavelength data were all taken at different epochs and their magnitudes even at the same wavelengths are variable as shown in Table~\ref{tab:phot_table}. 
However, since the photo-z derived from a hybrid template using SDSS is a large deviation from other estimates and with the poorest $\chi^2$, we think it is reasonable to exclude this redshift estimate. We therefore conclude that the redshift of \textit{WISE-J0301} is $z\sim 0.28-0.31$. The results of SED fits are summarized in Table~\ref{tab:photo_z_table}.


\begin{table*}[ht]
\tabletypesize{\scriptsize}
\centering
\begin{tabular}{ p{3cm}p{5cm}p{5cm}}
\hline
\hline
 & GALEX + SDSS + 2MASS  & GALEX + Pan-STARRS + 2MASS \\
Template & $z_{Phot}$ \,\,\, $\chi^{2}$ & $z_{Phot}$  \,\,\,  $\chi^{2}$ \\
\hline
Galaxy & 0.31 \,\,\, 11.13 & 0.28  \,\,\, \textbf{3.23} \\
AGN+Galaxy &  0.47  \,\,\, 18.35  & 0.28  \,\,\, 8.48\\
\hline
\end{tabular}
\caption{Redshift estimates derived from SED modeling of \textit{WISE-J0301}. SED fits to the \textit{GALEX} + SDSS + 2MASS and \textit{GALEX} + Pan-STARRS + 2MASS photometry are derived using 
galaxy templates derived from the FSPS model with EAZY method \citep{eazy} as well as mixed AGN/galaxy hybrid templates discussed in \citet{Salvato_AGN_temp_2009ApJ}. We note that Pan-STARRS photometry provides photo-$z$ estimates with the best $\chi^{2}$ values. }
\label{tab:photo_z_table}
\end{table*}

We note that our redshift estimate of \textbf{$z\sim 0.3$} 
is different from the photo-z, $z\sim$0.215, obtained by SDSS using KD-tree machine learning method \citep{sdss_dr12}. Unlike SDSS which only used $ugriz$, our estimates also take into account the \textit{GALEX} UV and 2MASS NIR photometry in addition to the optical photometry. It is well-known that the accuracy of photo-z depends on the quality of photometry and broad wavelength coverage with many photometric bands in observed galaxy SEDs \citep{Dahlen2010}. Thus, the photo-z estimate here would be more robust for our object with multi-wavelength photometry having high signal-to-noise ratios. 
 

Since the data seems to suggest that {\it WISE-J0301} is an obscured AGN, we fit the SED to a prototypical nearby obscured AGN, Mrk 231 ($z \sim 0.042$). At the best fit redshift of 0.28, the Mrk 231 template from  \citet{chary_elbaz_2001} shows a good fit to the observed MIR photometry with a $\chi^{2} =$ 5.5, after applying a scaling factor of 0.45. In Figure~\ref{fig:AGN_SED}, we plot the results of SED fit to the best-fitted galaxy template (green) and Mrk0231 template (black) with over-plotted observed photometry from \textit{GALEX}, PanSTARRS, 2MASS, \textit{WISE}, and NVSS.
The EAZY and Mrk 231 fits are used to estimate the physical parameters of the host galaxy which are tabulated in Table~\ref{tab:param_table}. We note that due to the variability in the optical photometry in the reddest bands, the derived parameters are uncertain at the 30\% level. In Section~\ref{sec:discussion}, we discuss our interpretation of these results.

\begin{figure*}[ht!]
\includegraphics[trim=0.0in 0.0in 0.in 0.in, clip=true, width=1.0\textwidth]{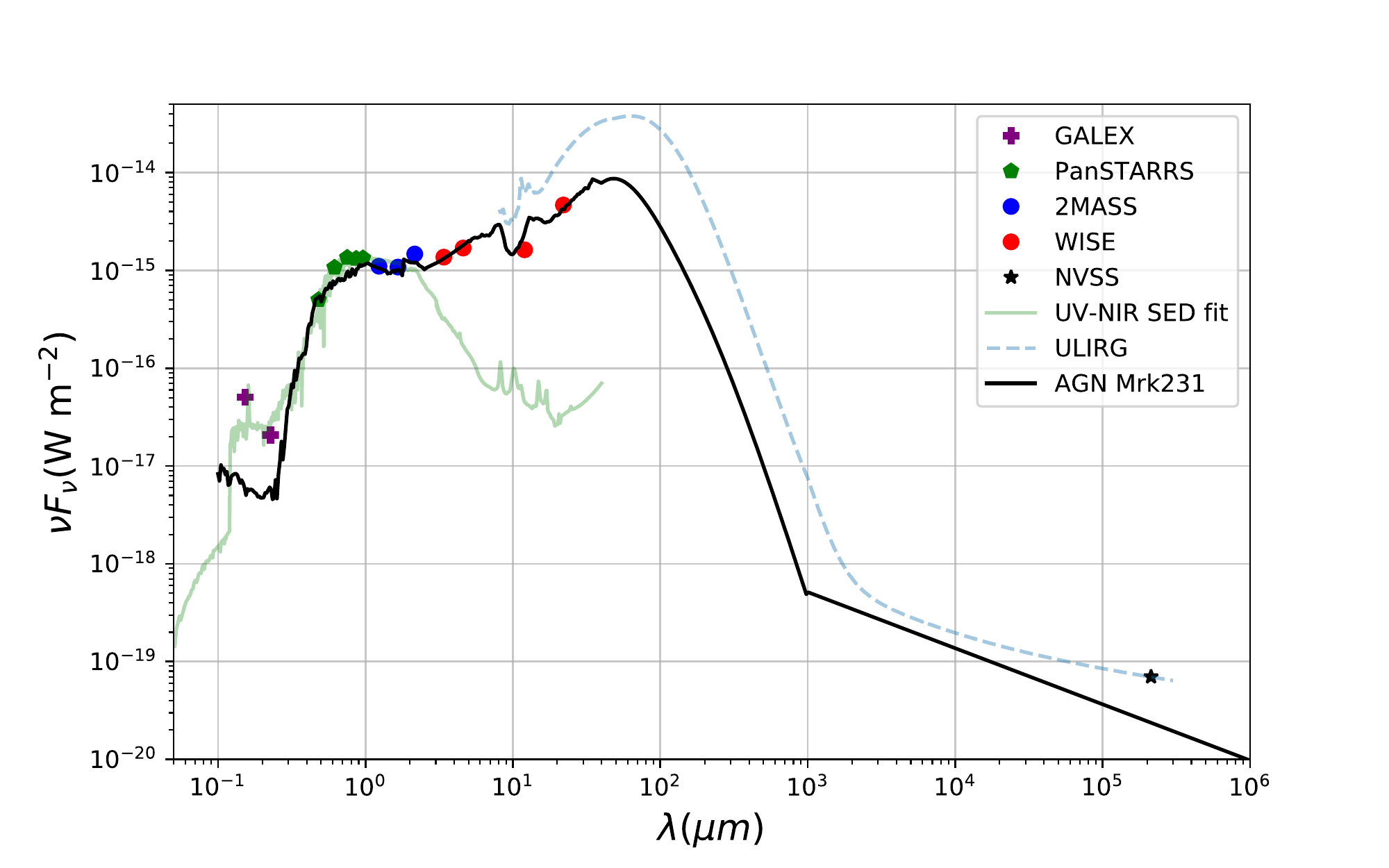}
\caption{The figure shows the observed time-averaged photometry of \textit{WISE-J0301} at observed-frame wavelengths from \textit{GALEX} (purple), PanSTARRS (green), 2MASS (blue), \textit{WISE} (red), and NVSS (black) plotted as data points. We show the model stellar population fit (green curve) to the UV-NIR photometry of the host galaxy. We fit the template of Mrk231 (black curve) to the MIR photometry to estimate bolometric IR luminosity. Dusty star-forming template from \citealt{chary_elbaz_2001} (dashed curve) scaled to fit radio photometry overshoots the observed MIR photometry. Physical parameters derived from these fits are summarized in Table~\ref{tab:param_table}.}
\end{figure*}
\label{fig:AGN_SED}

\section{results and analysis} \label{sec:discussion}
\subsection{WISE-J0301 and Mrk231: Similarities and differences} \label{sec:discussion_mrk231}
It is important to point out the key similarities as well as the differences between \textit{WISE-J0301} and Mrk231. While the Mrk231 mid- and far-infrared SED can be scaled to fit the observed photometry of \textit{WISE-J0301} extremely well, our extracted \textit{WISE} photometry of Mrk231 does not show long-term MIR variability like \textit{WISE-J0301} with a \textit{W1} and \textit{W2} brightness of 10.1 and 9.5 mags (AB), respectively. Mrk231 does show extreme variability at radio frequencies (\citealt{Mrk231_radio_2013}) with a range in 20 GHz flux densities of 75-275 mJy. We currently do not have any multi-epoch radio observations for \textit{WISE-J0301} to ascertain its radio variability. We however note that the radio to mid-infrared flux density ratio of \textit{WISE-J0301} is higher by a factor of $\sim$2.3 compared to Mrk231
as shown in Figure \ref{fig:AGN_SED}, suggesting an additional component of radio emission, likely from the AGN. 


\subsection{Optical-IR light curves} \label{sec:discussion_lc}
The evidence of MIR variability is clear in the \textit{WISE} \textit{W1} and \textit{W2} light curves (Figure~\ref{fig:wise_lc}). A $\sim$0.9 magnitude peak to peak variation, roughly corresponding to a flux ratio of $\sim$2, is observed between MJD-55625 to MJD-57425. However, extrapolating the light curve trend in the empty region between MJD-55625 to MJD-56675, we expect \textit{WISE-J0301} to be faintest around MJD-56025, which is not dissimilar from the minimum in the $y-$band light curve. 
The MIR and optical variability combined with the radio luminosity of the source, which exceeds 10$^{24}$\,W\,Hz$^{-1}$, is a definitive indicator of AGN activity at the center of \textit{WISE-J0301}. 



Rotating radio jet models have been successful in explaining the long-term variability ($\sim$10 years) of extragalactic radio sources in \textit{Planck} compact source catalog \citep{Planck_CScat_var_2013}. We made an attempt to model the variability of \textit{WISE-J0301} using rotating radio-jet model. Although the radio-jet model fits the light curve data prior to MJD-58025 very well, the latest observations from \textit{NEOWISE} and \textit{Spitzer} are inconsistent with the model. Further multi-epoch observations from \textit{ Spitzer} will provide more clarity on the variability and origin of emission as shown in the Figure~\ref{fig:wise_lc}.

It appears more likely that the long-term variability of {\it WISE-J0301} is simply due to variations in the accretion rate onto the black hole, a 
hypothesis which is a common scenario adopted for variable AGN \citep{peterson_1997}. Due to
obscuration by dust in the host galaxy, possibly in the form of a torus, only the red, relatively unextincted wavelengths of this emission are detected. The dust in the torus
absorbs the short-wavelength emission and re-emits it at mid- and far-infrared
wavelengths. The hot dust, at temperatures of around $\sim$700\,K is the
component that is seen in the WISE bands. This scenario would show a lag in the
light curves between the optical and mid-infrared bands depending on the light
travel time from the AGN to the surrounding torus. Future observations with {\it Spitzer} and data releases from PanSTARRS will be able to measure this time delay and assess the validity of this hypothesis.

\subsection{Spectral energy distribution}\label{sec:discussion_sed}
We have shown that the UV-NIR photometry of \textit{WISE-J0301}, obtained from \textit{GALEX}, PanSTARRS, and 2MASS, is well fit by stellar populations continuum model (solid green curve in Figure 3). This is used to place constraints on the stellar mass and star formation rate (SFR) of the galaxy 
We estimate the stellar mass and unobscured SFR from SED fits to be $\sim$~1.9 $\times$ 10$^{11}$ M$_{\odot}$, and 0.71 M$_{\odot}$ year$^{-1}$, respectively. 

We also have shown that a Mrk 231 template shows a reasonable fit to the
overall SED of this source while noting that dusty starburst templates which have the
standard radio-IR correlation \citep[dashed line in Figure 3, e.g.][]{chary_elbaz_2001} violate the 
WISE 12 and 22\,$\mu$m photometry. This indicates that the brightness of {\it WISE-J0301} in the radio radio is unlikely to be due to star formation and is dominated by AGN emission.


As mentioned below, a scaling factor of $\sim$0.45 is applied for the Mrk231 template to fit the MIR luminosity of \textit{WISE-J0301}. This scaling provides us with an estimate of flux densities at FIR wavelengths for \textit{WISE-J0301} where we currently do not have any observations. We integrate the estimated IR flux from $8-1000\mu$m to obtain an 9.0 $\times$ 10$^{11}$ L$_{\odot}$ $\leq $ L$_{8-1000\mu m}$ (L$_{IR}$) $\leq $ 1.5 $\times$ 10$^{12}$ L$_{\odot}$ which suggest that the host galaxy is likely a ULIRG to within the uncertainties. More recent studies of Mrk231 and similar low redshift AGN/ULIRG \citep{veilleux_ulirg_2009,Weedman_mrk231_2005}, estimate the AGN contribution to the bolometric luminosity to be $\sim$70-80\%, consistent with the weakness of the PAH features in its mid-infrared SED. Assuming a $\sim$75\% contribution from AGN and the remaining $\sim$25\% from star formation, the SFR is estimated to be $\sim$ 70 M$_{\odot}$/yr ($\sim$100 times the SFR estimated from UV photometry). We examined the Infrared Astronomical Satellite (IRAS; \citet{IRAS_1984}) data on \textit{WISE-J0301}, using the SCANPI data service 
at IRSA, and found no detection in any of the 4 IRAS bands.\footnote{\url{https://irsa.ipac.caltech.edu/applications/Scanpi/}} The 
baseline noise combined between 60 and 100 $\mu$m implies a 1-$\sigma$ on FIR 
flux of $\sim$7.23 10$^{-15} W m^{-2}$. This gives a 1-$\sigma$ equivalent in SFR of 67 M$_{\odot}/yr$ which broadly is consistent with the SFR estimate from scaled Mrk231 template. 
Overall, these estimates lead us to conclude that \textit{WISE-J0301} is possibly a ULIRG with strong AGN activity and some star-formation in it.

It has long been believed that ULIRGs, at low redshifts, are driven by mergers between gas-rich galaxies, where the interaction triggers dust-enveloped starburst and AGN activity \citep{gao_and_solomon_2004}. We do not see any obvious evidence of merger activity in the images obtained in PanSTARRS and \textit{WISE} \textit{W1} bands (Figure~\ref{fig:ps_wise_img}). This suggests the possibility that the host galaxy is at a very late stage of merger hence showing a smooth extended shape at the resolution of the data.


\subsection{Constraints on Black Hole Mass}\label{sec:discussion_agn}
We next constrain the mass of the BH at the center of \textit{WISE-J0301}.
Given the similarity between the SED of this object and Mrk 231, and the fact
that the SED is dominated by AGN emission, one estimate comes from the direct scaling factor.
The black hole at the center of AGN Mrk231 has an estimated mass of $\sim$2.3 $\times$ 10$^{8}$ $M_{\odot}$ \citep{Leighly_2014}. As stated before, we estimate a scaling of $\sim$0.45 for AGN Mrk231 to fit the observed MIR luminosity of \textit{WISE-J0301}. If we assume that the FIR SED of this object is similar to that of Mrk231, we derive an upper limit on the mass of the BH at the center of \textit{WISE-J0301} to be $\sim$1.0 $\times$ 10$^{8}$ $M_{\odot}$.

We also extracted the X-ray brightness of \textit{WISE-J0301} from archival \textit{Chandra} X-ray data. We find L$_{X-ray}$ $\sim$ 2 $\times$ 10$^{43}$ ergs s$^{-1}$. In order to correct for absorption to the X-ray brightness, we need to use the MIR - X-ray luminosity correlation demonstrated by \citet{Lutz_2004}. We estimate a 6 $\mu$m flux density from only the variable component of 
\textit{WISE} \textit{W1}, and \textit{W2} fluxes (stellar component subtracted) using a simple power-law shown in Equation~\ref{eqn:p_law}. We estimate L$_{6\mu m}$ $\sim$ 6 $\times$ 10$^{44}$ ergs s$^{-1}$ which indicates an absorption-correction of $\sim$ 5 to the measured X-ray flux. Assuming L$_{X-ray}$ $\sim$ 0.1 $\times$ L$_{Bol}$, consistent with the measured 6$\mu$m luminosity, the absorption-corrected X-ray luminosity of L$_{X-ray}$ $\sim$ 1 $\times$ 10$^{44}$ ergs s$^{-1}$ places a lower limit on the BH mass of 1.0 $\times$ 10$^{7}$ $M_{\odot}$, assuming Eddington limited accretion;


\begin{eqnarray}
L_{Edd} = 3.2 \times 10^{4} \times \frac{M_{BH}}{M_{\odot}} \times L_{\odot}, \label{eqn:fir_sfr}
\end{eqnarray}
where M$_{BH}$ is the mass of the central BH. 
Naturally, if the accretion rate were sub-Eddington, the resultant black hole mass would be much higher. Furthermore,
if \textit{WISE-J0301} turns out to be brighter than Mrk231 in FIR, it could either imply a higher accretion rate or a higher star-formation rate than in Mrk231.

It is interesting to analyze the properties of \textit{WISE-J0301} in the context of BH to stellar mass relations of AGN in the low-z universe \citep{reines_volonteri_2015}. For the estimated stellar mass of $\sim$2 $\times$10$^{11}$ M$_{\odot}$, a BH mass of $\sim$ 10$^7$ M$_{\odot}$ is substantially below the canonical black hole to stellar mass relationship presented in \citet{mcConnell_2013}. However, a BH mass of 
10$^{8}$ M$_{\odot}$ is consistent with the relationship found for broad line AGN, arguing in favor of the higher value and suggesting that the FIR SED of this source may indeed be similar to Mrk231.

\begin{table}[ht]
\tabletypesize{\scriptsize}
\centering
\begin{tabular}{llll}
\hline
\hline
SFR$_{UV}$ & 0.71 $M_{\odot}$/yr \\
Stellar Mass & $1.9 \times 10^{11}$ $M_{\odot}$ \\
L$_{UV-NIR}$   &  $0.8 \times 10^{11}$ - $1.7 \times 10^{11}$ $L_{\odot}$ \\
SFR$_{8-1000\mu m}$(SFR$_{IR})$     & $\sim$70 $M_{\odot}$/yr\\
L$_{8-1000\mu m}$ (L$_{IR}$) & $9.0 \times 10^{11}$ - $1.5 \times 10^{12}$ $L_{\odot}$ \\
BH MASS & $\sim$ $10^7 - 10^{8}$ $M_{\odot}$\\
\hline
\end{tabular}
\caption{Physical parameters derived from SED modeling of \textit{WISE-J0301}. SED fit to the UV-NIR photometry is derived using the optimized EAZY template set discussed in Section 4. The FIR luminosities and star-formation rates are derived by scaling the Mrk231 template from \citet{chary_elbaz_2001}.}
\label{tab:param_table}
\end{table}

\section{Conclusions and future works} \label{sec:results}
\textit{WISE} \textit{J030654.88+010833.6} (\textit{WISE-J0301}) at redshift, 0.28 $\leq$ $z\leq$0.31, is one of the few known MIR variable, radio detected AGN hosted in a ULIRG which shows optical variability at redder wavelengths. The model SED fit to the UV-MIR photometry combined with AGN Mrk231 template fit to MIR photometry indicate that \textit{WISE-J0301} could possibly be a composite AGN/starburst ULIRG in a phase where high star formation (SFR $\sim$ 70 M$_{\odot}$ year$^{-1}$) is occurring. The black hole to stellar mass ratio of \textit{WISE-J0301} also appears to be consistent with that of broad line AGN in the local universe. Our estimates of the SFR$_{IR}$ and BH mass of \textit{WISE-J0301} are highly uncertain and based on scaling the Mrk231 SED template to the MIR photometry. We currently do not have any publicly available spectra or FIR photometry for \textit{WISE-J0301} which would provide more robust estimates of these parameters. The optical and MIR images do not show any clear evidence of merger activity which is popularly believed to be the driving force behind ULIRG at lower redshifts. This could possibly be an indicator of the fact that \textit{WISE-J0301} is at a very late stage of a merger. The long-term MIR variability, combined with the red optical colors of the emission, seems to suggest that the variability is because of variations in the accretion rate of the black hole which
is partly obscured by dust. The dust thermally reprocesses some of the luminosity emitted by the accretion disk; thus the variation in mid-infrared luminosity should
lag that seen in the optical bands by the light travel time across a dusty torus. 

In the future, we plan to obtain FIR measurements for \textit{WISE-J0301}, which is necessary to disentangle the contribution of the AGN and star formation to the bolometric luminosity. Further, we currently have a Spitzer space telescope DDT approved program to extend the MIR light curves by $\sim$6 months, post \textit{NEOWISE} (June 2019), which when combined with future PanSTARRS data
will allow us to further investigate the origins of MIR variability. \textit{WISE-J0301} provides us with a rare opportunity to put important constraints on the growth of SMBH and their connection to galaxy evolution at late cosmic times.

\section{ACKNOWLEDGMENTS}
This work was supported by Joint Survey Processing (JSP) at IPAC/Caltech which is aimed at combined analysis of Euclid, LSST, and WFIRST. Funding for JSP has been provided by NASA grant 80NM0018F0803.

This publication makes use of data products from the Wide-field Infrared Survey Explorer, which is a joint project of the University of California, Los Angeles, and the Jet Propulsion Laboratory/California Institute of Technology, funded by the National Aeronautics and Space Administration.

This research has also made use of the NASA/IPAC Extragalactic Database (NED) which is operated by JPL/Caltech, under contract with NASA. Ned Wright’s Cosmology Calculator was also used in preparing this paper.

We thank the {\it Spitzer} Science Center Director and observation support team for an award of Director's Discretionary Time for this project. 

Funding for the Sloan Digital Sky Survey IV has been provided by the Alfred P. Sloan Foundation, the U.S. Department of Energy Office of Science, and the Participating Institutions. SDSS acknowledges support and resources from the Center for High-Performance Computing at the University of Utah. The SDSS web site is www.sdss.org.

SDSS is managed by the Astrophysical Research Consortium for the Participating Institutions of the SDSS Collaboration including the Brazilian Participation Group, the Carnegie Institution for Science, Carnegie Mellon University, the Chilean Participation Group, the French Participation Group, Harvard-Smithsonian Center for Astrophysics, Instituto de Astrofísica de Canarias, The Johns Hopkins University, Kavli Institute for the Physics and Mathematics of the Universe (IPMU) / University of Tokyo, the Korean Participation Group, Lawrence Berkeley National Laboratory, Leibniz Institut für Astrophysik Potsdam (AIP), Max-Planck-Institut für Astronomie (MPIA Heidelberg), Max-Planck-Institut für Astrophysik (MPA Garching), Max-Planck-Institut für Extraterrestrische Physik (MPE), National Astronomical Observatories of China, New Mexico State University, New York University, University of Notre Dame, Observatório Nacional / MCTI, The Ohio State University, Pennsylvania State University, Shanghai Astronomical Observatory, United Kingdom Participation Group, Universidad Nacional Autónoma de México, University of Arizona, University of Colorado Boulder, University of Oxford, University of Portsmouth, University of Utah, University of Virginia, University of Washington, University of Wisconsin, Vanderbilt University, and Yale University.

The Pan-STARRS1 Surveys (PS1) have been made possible through contributions of the Institute for Astronomy, the University of Hawaii, the Pan-STARRS Project Office, the Max-Planck Society and its participating institutes, the Max Planck Institute for Astronomy, Heidelberg and the Max Planck Institute for Extraterrestrial Physics, Garching, The Johns Hopkins University, Durham University, the University of Edinburgh, Queen's University Belfast, the Harvard-Smithsonian Center for Astrophysics, the Las Cumbres Observatory Global Telescope Network Incorporated, the National Central University of Taiwan, the Space Telescope Science Institute, the National Aeronautics and Space Administration under Grant No. NNX08AR22G issued through the Planetary Science Division of the NASA Science Mission Directorate, the National Science Foundation under Grant No. AST-1238877, the University of Maryland, and Eotvos Lorand University (ELTE).

The Catalina Real-time Transient survey is funded by the National Aeronautics and Space Administration under Grant No. NNG05GF22G issued through the Science
Mission Directorate Near-Earth Objects Observations Program.  The CRTS
survey is supported by the U.S.~National Science Foundation under
grants AST-0909182.


\bibliographystyle{apj}
\bibliography{AGN_ref}\scriptsize

\end{document}